\documentclass[a4paper,11pt,oneside,headsepline,twocolumn]{scrartcl}

\usepackage{ifdraft}

\usepackage[utf8]{inputenc}
\usepackage[T1]{fontenc}

\usepackage{lmodern}
\usepackage{fourier}
\usepackage{amssymb, amsfonts}

\usepackage{mathrsfs} % script characters (\mathscr)
\usepackage{dsfont}
\usepackage[usenames,dvipsnames]{xcolor}%, colortbl}
\usepackage{lettrine}
\usepackage{url}

\usepackage{stmaryrd}
\usepackage[english]{babel}

\usepackage[draft=false]{scrlayer-scrpage}
\usepackage{wrapfig}
\usepackage[compact]{titlesec}
\usepackage{footmisc}
\usepackage{needspace}

\usepackage{amsmath}
\usepackage[thmmarks, amsmath, amsthm]{ntheorem}

\usepackage[style=alphabetic,url=false,sortcites=true,maxbibnames=99,backend=bibtex]{biblatex}
\bibliography{bibliography.bib}

\ifdraft{{}}
  {\usepackage[pdftex,
               pdfborder={0 0 0}, colorlinks=true,
               linkcolor=BrickRed, citecolor=ForestGreen, urlcolor=RoyalBlue]{hyperref}}

\ifdraft{\usepackage[colorinlistoftodos]{todonotes}}{}

\usepackage{booktabs}
\usepackage{caption}
\usepackage{csquotes}
\usepackage[inline]{enumitem}
\usepackage{float}
\usepackage{graphicx}
\usepackage[final]{listings}
\usepackage{numprint}
\ifdraft{\date{\today}}{\date{}}

\KOMAoptions{DIV=13}

\clearscrheadings
\automark[section]{subsection}
\renewcommand{\sectionmark}[1]{\markright{\thesection\;#1}}
\renewcommand{\subsectionmark}[1]{}

\lohead{{\small \headertitle \,---\, \rightmark}}
\rohead{{\small \headerauthors}}

\newenvironment{plainfootnotes}{
  \deffootnote[0em]{0em}{0em}{}
}{
  \deffootnote[1em]{1.5em}{1em}{\textsuperscript{\thefootnotemark}}
}

\newif\ifsubsectionstylePrefixParagraph
\newif\ifsubsectionstyleRunin

\subsectionstylePrefixParagraphtrue
\subsectionstyleRunintrue

\titleformat{\section}[hang]
{\Large\sffamily\bfseries}
{\thesection\hspace{0.25em}{}}{0.25em}{}

\ifsubsectionstyleRunin
\ifsubsectionstylePrefixParagraph

\titleformat{\subsection}[runin]
{\normalfont\bfseries}
{\S\hspace{.1em}\thesubsection}{0.35em}{}[.\hspace*{.5em}]

\else

\titleformat{\subsection}[runin]
{\normalfont\bfseries}
{\thesubsection\hspace{.1em}|}{0.25em}{}[.\hspace*{.5em}]

\fi
\else
\ifsubsectionstylePrefixParagraph

\titleformat{\subsection}[wrap]
{\normalfont\bfseries\selectfont\filright}
{\S\thesubsection}{.35em}{}
\titlespacing{\subsection}
{16pc}{1.5ex plus .1ex minus .2ex}{1pc}

\else

\titleformat{\subsection}[wrap]
{\normalfont\bfseries\selectfont\filright}
{\thesubsection\hspace{.1em}|}{.25em}{}
\titlespacing{\subsection}
{16pc}{1.5ex plus .1ex minus .2ex}{1pc}

\fi
\fi

\ifsubsectionstylePrefixParagraph

\titleformat{\subsubsection}[runin]
{\normalfont\bfseries}
{\S\hspace{.1em}\thesubsubsection}{0.35em}{}[.]

\else

\titleformat{\subsubsection}[runin]
{\normalfont\bfseries}
{\thesubsubsection\hspace{.1em}|}{0.25em}{}[.]
  
\fi

\titleformat{\paragraph}[runin]
{\normalfont\itshape}
{\S\hspace{.1em}\theparagraph}{0.35em}{}[.]

\AtEveryBibitem{\clearfield{isbn}}
\AtEveryBibitem{\clearlist{language}}
\AtEveryBibitem{\clearfield{pages}}

\newenvironment{enumeratearabic}{
\begin{enumerate}[label=(\arabic*), leftmargin=0pt,labelindent=2em,itemindent=!]
}{
\end{enumerate}
}

\newenvironment{enumeratearabic*}{
\begin{enumerate*}[label=(\arabic*)] %%, leftmargin=0pt,labelindent=2em,itemindent=!]
}{
\end{enumerate*}
}

\newenvironment{enumerateroman*}{
\begin{enumerate*}[label=(\roman*)] %%, leftmargin=0pt,labelindent=2em,itemindent=!]
}{
\end{enumerate*}
}

\numberwithin{equation}{section}

\newtheorem{theoremcounter}{theoremcounter}[section]

\theoremstyle{plain}

\theoremnumbering{Roman}

\theoremstyle{definition}

\newtheorem{definition}[theoremcounter]{Definition}

\theoremstyle{remark}

\newtheorem{example}[theoremcounter]{Example}

\newtheorem{remark}[theoremcounter]{Remark}

\newtheorem*{mainremark}{Remark}

\newtheorem*{remarkcomputation}{Computation}

\ifdraft{
 
}{
 
}

\newcommand{\tx}{\ensuremath{\text}}

\newcommand{\tbf}{\bfseries}

\newcommand{\thdash}{\nbd th}

\newcommand{\nbd}{\nobreakdash-\hspace{0pt}}

\newcommand{\bboard}{\ensuremath{\mathbb}}

\newcommand{\bbF}{\ensuremath{\bboard F}}

\newcommand{\rmb}{\ensuremath{\mathrm{b}}}

\newcommand{\rml}{\ensuremath{\mathrm{l}}}

\newcommand{\rmr}{\ensuremath{\mathrm{r}}}

\newcommand{\rmt}{\ensuremath{\mathrm{t}}}

\newcommand{\ZZ}{\ensuremath{\mathbb{Z}}}
\newcommand{\QQ}{\ensuremath{\mathbb{Q}}}
\newcommand{\RR}{\ensuremath{\mathbb{R}}}
\newcommand{\CC}{\ensuremath{\mathbb{C}}}

\newenvironment{psmatrix}{\left(\begin{smallmatrix}}{\end{smallmatrix}\right)}

\newcommand{\Mat}[1]{\ensuremath{\mathrm{Mat}_{#1}}}
\newcommand{\T}{\ensuremath{\rmt}}
\newcommand{\rT}{\ensuremath{\,{}^\T\!}}

\newcommand{\R}{\mathcal{R}}
\newcommand{\K}{\mathcal{K}}

\newcommand{\headertitle}{{\normalfont%
  HLinear
}}
\newcommand{\headerauthors}{
  A.~Ghitza,
  M.~Raum
}

\begin{document}

\binoppenalty=9999
\relpenalty=9999

\begin{plainfootnotes}
\begin{flushleft}
{\fontfamily{lms}\sffamily
  \hspace{20pt}{\huge%
  HLinear
  % \\\hspace{20pt}{\huge%
  % %% TITLE LINE 2
  }
}
\\{\fontfamily{lms}\sffamily
  \hspace{20pt}%
  Exact Dense Linear Algebra in Haskell
}
\\[.6em]\hspace{20pt}{\large%
  Alexandru Ghitza%
  \footnote{The first author was partially supported by ARC Discovery Grant DP120101942.}
  and}\\\hspace{20pt}{\large%
  Martin Raum%
  \footnote{The second author was partially supported by Vetenskapsr\aa det Grant 2015-04139.}
}
\\[1.2em]
\end{flushleft}
\end{plainfootnotes}

\thispagestyle{scrplain}

%%%%%%%%%%%%%%%%%%%%%%%%%%%%%%%%%%%%%%%%%%%%%%%%%%
%%% ABSTRACT

{\small
\noindent
{\tbf Abstract:}
We present an implementation in the functional programming language Haskell of the PLE decomposition of matrices over division rings. Our benchmarks indicate that it is competitive with the C-based implementation provided in Flint. Describing the guiding principles of our work, we introduce the reader to basic ideas from high-performance functional programming.
\\[.35em]
%
%%
% \textsf{\textbf{%
%   KEYWORD 1%
% \hspace{0.3em}{\tiny$\blacksquare$}\hspace{0.3em}%
%   KEYWORD 2%
% \hspace{0.3em}{\tiny$\blacksquare$}\hspace{0.3em}%
%   KEYWORD 3%
% }}
% \\[0.15em]
% %
% \noindent
% \textsf{\textbf{%
%   MSC Primary:
%   INSERT%
% \hspace{0.3em}{\tiny$\blacksquare$}\hspace{0.3em}%
%   MSC Secondary:
%   INSERT
% }}
}

%%%%%%%%%%%%%%%%%%%%%%%%%%%%%%%%%%%%%%%%%%%%%%%%%%
%%% TABLE OF CONTENTS

% \vspace{-1.5em}
% \renewcommand{\contentsname}{}
% \setcounter{tocdepth}{2}
% \tableofcontents
% \vspace{1.5em}

%%%%%%%%%%%%%%%%%%%%%%%%%%%%%%%%%%%%%%%%%%%%%%%%%%
%%% INTRODUCTION

\Needspace*{4em}
\addcontentsline{toc}{section}{Introduction}
\markright{Introduction}
\lettrine[lines=2,nindent=.2em]{\tbf L}{inear} algebra pervades modern algorithms. Today's multitude of applications of this field has spawned an equal multitude of refinements of it. Dense (as opposed to sparse) linear algebra refers to the computation with matrices or vectors with few expected zero entries. Exact linear algebra (as opposed to approximate linear algebra~\cite{kreinovich-lakeyev-noskov-1996} and numerical linear algebra) refers to computation admitting no
approximation error. Need for exact dense linear algebra arises from, among others, cryptography, compression, and problems in pure mathematics. It is the backbone of symbolic and algebraic-geometric computation facilities.

Major open-source implementations of exact dense linear algebra are available within LinBox~\cite{linbox-1.3.2} and Flint~\cite{flint-2.5.2} (Fast LIbrary for Number Theory). LinBox has functionality for computing: solutions to linear equations, invariants of linear operators, and various canonical forms of matrices. The focus is on computation over finite fields and the integers, and extends to rationals via a technique called rational reconstruction. The library started out with an emphasis on black box algorithms~\cite{kaltofen-trager-1990}, but now also implements the typical algorithms for dense matrices.
Flint offers functionality similar to LinBox, but uses a different way of implementing it. It is mostly based on classical and Strassen approaches~\cite{strassen-1969}.

Questions about vector spaces and systems of linear equations are mostly addressed via matrix factorizations. One instance of matrix factorization is the PLE decomposition of a matrix~$M$, which in particular provides an echelon form~$E$. In this work, we present an implementation \emph{HLinear} of the PLE decomposition in the functional programming language Haskell. It is competitive with Flint and in some cases outperforms it (see the Performance section). On the other hand, it enjoys typical benefits of programs written in functional languages. For instance, it opens doors to formal verification and painless distributed computation.

The HLinear code is available in the public repository
\begin{center}
  \url{https://github.com/martinra/hlinear/tree/paper-toms}
\end{center}
A build script that installs locally the correct version of Flint, and checks out the Haskell code is available at
\begin{center}
  \url{http://www.raum-brothers.eu/martin/index.html#downloads}
\end{center}

%%%%%%%%%%%%%%%%%%%%%%%%%%%%%%%%%%%%%%%%%%%%%%%%%%
%%% PAPER BODY

\section{Background}

\subsection{Motivation}

The need for an efficient PLE~decomposition grew out of the second author's project to compute with (Siegel) modular forms---cf.\ the last section of~\cite{bruinier-raum-2014}. Matrices arising from this application are comparatively large with 10,000 up to several 100,000 rows. To complicate matters, they have entries over number fields\footnote{We should point out that, while HLinear works well with very general coefficients, matrices of 100,000 rows are currently out of reach, as can be seen in the Performance section.}. On the plus side, algebraic-geometric methods show that these matrices have PLE~decompositions with rather small denominators. The urgent need for parallelization\footnote{We note that there is work in progress for parallelizing certain algorithms in LinBox~\cite{dumas-gautier-pernet-saunders-2010}.} and
distributed computing in conjunction with the authors' desire to formally verify as much of their future computation as possible, rendered impossible the usage of available implementations. The verification requirement, specifically, suggested use of a functional programming language.

\subsection{PLE decomposition}
\label{ssec:ple-decomposition}

Let $\R$ be a (unital) ring.
Given a matrix $M \in\nobreak \Mat{m,n}(\R)$, we say that $M = PLE$ is a PLE~decomposition of~$M$, if $P$ is a permutation matrix, $L$ is a lower triangular matrix with diagonal entries~$1$, and $E$ is a matrix in echelon form. Jeannerod, Pernet, and Storjohann~\cite{jeannerod-pernet-storjohann-2013} explain the PLE and related decompositions in the context of rank-profiles. As an example of PLE~decomposition, we record that a $4 \times 6$ matrix could allow for the following factorization:
\begin{gather*}
  P
  \begin{psmatrix}
    1 \\
    \ast & 1 \\
    \ast & \ast & 1 \\
    \ast & \ast & \ast & 1 \\
  \end{psmatrix}
  \begin{psmatrix}
    \ast' & \ast & \ast & \ast  & \ast  & \ast \\
          &      &      & \ast' & \ast  & \ast \\
          &      &      &       & \ast' & \ast \\
          &      &      &       &       &      \\
  \end{psmatrix}
\tx{,}
\end{gather*}
where $\ast$ is an arbitrary entry from $\R$ and $\ast'$ is a non-zero entry.

If $\R = \K$ is a field (or a division ring) then it is possible to pass to normalized echelon forms. In this case, we allow for arbitrary non-zero entries on the diagonal of $L$, and in exchange demand that the pivot entries of $E$ be~$1$.
\begin{gather*}
  P
  \begin{psmatrix}
    \ast' \\
    \ast & \ast' \\
    \ast & \ast & \ast' \\
    \ast & \ast & \ast & \ast' \\
  \end{psmatrix}
  \begin{psmatrix}
    1 & \ast & \ast & \ast & \ast & \ast \\
      &      &      & 1    & \ast & \ast \\
      &      &      &      & 1    & \ast \\
      &      &      &      &      &      \\
  \end{psmatrix}
\tx{.}
\end{gather*}
Slightly ambiguously, this decomposition is also called (normalized) PLE~decomposition. To obtain a normal form of $M$ with respect to the action of invertible matrices from the right, one may proceed to the reduced echelon normal form by applying to $E$ an upper triangular matrix $U$ with diagonal entries~$1$. We thus obtain the PLUE~decomposition associated with the previous example:
\begin{gather*}
  P
  \begin{psmatrix}
    \ast' \\
    \ast & \ast' \\
    \ast & \ast & \ast' \\
    \ast & \ast & \ast & \ast' \\
  \end{psmatrix}
  \begin{psmatrix}
  1 & \ast & \ast &   \\
    & 1    & \ast &   \\
    &      & 1    &   \\
    &      &      & 1 \\
  \end{psmatrix}
  \begin{psmatrix}
    1 & \ast & \ast &   &   & \ast \\
      &      &      & 1 &   & \ast \\
      &      &      &   & 1 & \ast \\
      &      &      &   &   &      \\
  \end{psmatrix}
\tx{.}
\end{gather*}

Gaussian elimination is the most classical algorithm to compute PLE decompositions. It proceeds by iterating through columns:
\begin{enumeratearabic*}
\item picking a non-zero element in the current column, if possible;
\item permuting the corresponding row to the top unprocessed position;
\item normalizing that row;
\item eliminating entries in the current column below that row.
\end{enumeratearabic*}
Despite its age, this algorithm continues to be the fundamental building block in the computation of PLE decompositions of modest-sized matrices.

One alternative to Gaussian elimination is the slice PLE decomposition (see~\cite{bunch-hopcroft-1974} and, for example,~\cite{albrecht-ple} for a recent application), which is a hierarchical approach. Splitting up $M$ into column slices $M = \begin{psmatrix} M_0 & \cdots & M_{r-1} \end{psmatrix}$, one computes $M_0 = P_0 L_0 E_0$ and sets $M'_i = L_0^{-1} P_0^{-1} M_i$ for $i \ge 1$. Then we decompose $M'_i = \rT \begin{psmatrix} \rT E_i & \rT M''_i \end{psmatrix}$ row-wise, where the number of rows of $E_i$ equals the rank of~$E$. Setting $M'' = \begin{psmatrix} M''_1 & \cdots & M''_{r-1} \end{psmatrix}$ allows to find $M'' = P'' L'' E''$ by recursion, and thus build the PLE decomposition of~$M$ by rearranging
\begin{gather*}
  M
=
  P_0 L_0 E_0 \cdot
  \begin{psmatrix}
    E_1 \; \cdots \; E_{r-1} \\
      P'' L'' E''
  \end{psmatrix}
\tx{.}
\end{gather*}
This approach and its iterative counterpart have been implemented in M4RI~\cite{m4ri} and in a not-yet-released version of LinBox.

Given that LinBox employs rational reconstruction, we revisit multi-modular linear algebra. The key observation is that the decomposition $M = PLE$ for $M \in \Mat{m,n}(\QQ)$ can be reconstructed from its reductions modulo a large enough integer $N$, coprime to the denominators of $M$, $L$, and $E$.
In practice $N$ can be chosen to be a product of distinct word-sized primes, which leads to vastly reduced coefficient size.
Of course, this comes at the expense of additional reconstruction steps.

\subsection{Implementations of PLE decompositions}
\label{ssec:background-implementations}

From the three approaches to linear algebra presented above, we see that dense linear algebra requires optimization on at least two scales. First of all, it involves frequent additions and multiplications in the coefficient ring $\R$, and occasional divisions. Such operations are typically optimized at low level. For exact computation, this is addressed in libraries like GMP~\cite{gmp} and MPIR~\cite{mpir}. Second, the structural dependencies among the operations require
optimization at high level. They are traditionally met by studying the algorithms from a theoretical point of view. Modern compilers and libraries facilitate exploitation of fusion, dependency analysis, and even term rewriting.

Other aspects that receive increasing attention are reliability, security, and correctness. They are of considerable importance to cryptography and pure mathematical applications. Correctness is typically addressed by testing, which is supported by various frameworks available for major programming languages. Formal verification provides further reassurance, but it is hard to apply to today's most popular languages due to, for example, insufficient type systems.

We have mentioned the two implementations LinBox and Flint of exact dense linear algebra. LinBox is more established and aims to be a general-purpose library; it makes heavy use of rational reconstruction. Flint's linear algebra implementation is more recent and geared primarily toward the needs of number-theoretic computations, using classical algorithms. For work over the rationals, Flint mostly performs a little bit better than LinBox.
The existence and continued development of the two libraries have been very beneficial to users of computational linear algebra.

LinBox is written in C++, while Flint is written in C. We notice that Flint cannot rely on the compiler's ability to simplify structure at a larger scale. Indeed, it focuses on low-level optimizations, and deals with high-level optimizations by hand. LinBox can and does rely on templates for achieving a certain level of generality and structural optimization.

Both C and C++ have excellent properties when it comes to program overhead; LinBox and Flint make use of them. On the other hand, neither supports the programmer with optimization of large scale structures, testing, or verification. These are features of functional programming that could be profitably exploited for computational linear algebra.

\subsection{Haskell}

Haskell is a functional programming language, the most popular one after
OCaml. It is successful due to, among other things, the highly-developed and
aggressively optimizing Glasgow Haskell Compiler (GHC). While functional
programming languages suffer the reputation of being relatively slow, recent
progress on fusion~\cite{mainland-stream-fusion} and plenty of highly-developed libraries allowed for implementing the high-performance webserver
Warp~\cite{warp}, beating C code on some numerical
applications~\cite{mainland-stream-fusion}, and software being employed in the
financial industry\footnote{Examples include Barclays Capital~\cite{frankau},
Credit Suisse~\cite{credit-suisse}, Deutsche Bank~\cite{deutsche-bank}.}.
Haskell supports the developer by providing compositional code, term rewriting
rules, and a strong type system. As a result, code reusability, testing, and
verification (in connection with theorem provers such as Coq~\cite{coq} and
Isabelle/HOL~\cite{isabelle}) are superior to any other language encountered
in an industrial setting.

Haskell has important weak points: (a) it is infamous for its steep learning curve; (b) it does not have a type system as strong as, say, Agda~\cite{agda}; (c) it has several low-hanging fruits to be optimized in its parallelization and distributed computation libraries. Regardless of these imperfections, it currently appears as the best possible choice for functional (in the sense of functional programming) implementations.

We refer the reader who is unfamiliar with Haskell to~\cite{learning-haskell}, and illustrate code re\-usability in Haskell by a design pattern that we will encounter later. It is a common scheme to (i) decompose a data structure into a sequence of relatively simple data structures, and then (ii) recombine these simple structures. In Haskell, this is supported by unfolding and folding. The respective type signatures are\footnote{In some of the code listings, we had to violate Haskell's
indentation rules in order to make the lines fit. We also simplified certain definitions for clarity of exposition.}
\begin{lstlisting}[language=haskell,basicstyle=\small]
unfoldr :: (b -> Maybe (a, b)) -> b -> [a]
foldl :: (b -> a -> b) -> b -> [a] -> b
\end{lstlisting}
That is, unfolding is based on a function that decomposes an instance of a data structure \textsf{b}, if possible, into an easier piece of type~\textsf{a} and a remainder, which is again of type~\textsf{b}.

As a simple example, we formulate the extended Euclidean algorithm for non-negative integers (corresponding to the type~\textsf{Natural}) in terms of fold and unfold. Thinking of pairs $(a,b)$ as row vectors, a single reduction step in the Euclidean algorithm can be viewed as writing $\begin{psmatrix} a & b \end{psmatrix} = \begin{psmatrix} b & r \end{psmatrix}T$ for the matrix~$T=\begin{psmatrix}t & 1\\1 & 0\end{psmatrix}$, where $a=tb+r$ is the result of the division of $a$ by $b$. This reduction step is implemented as:
\begin{lstlisting}[language=haskell,basicstyle=\small]
reduce (_,0) = Nothing
reduce (0,b) = Just ((-1), (b,b))
reduce (a,b) = let (t,r) = quotRem a b
               in Just (t, (b,r))
\end{lstlisting}
From a pair $(a,b)$ we thus obtain a list $[T_1,\ldots,T_r]$ such that $\begin{psmatrix} a & b \end{psmatrix} = \begin{psmatrix} 1 & 0\end{psmatrix}T_1 \cdots T_r$. The matrix $\big( T_1 \cdots T_r \big)^{-1} = T_r^{-1} \cdots T_1^{-1}$ is then computed by folding via:
\begin{lstlisting}[language=haskell,basicstyle=\small]
mulinv (a,b,c,d) t = (b,a-t*b,d,c-t*d)
\end{lstlisting}
The extended Euclidean algorithm on a pair~$(a,b)$ can thus be cleanly written as
\begin{lstlisting}[language=haskell,basicstyle=\small]
let (x,_,y,_) = foldl mulinv (1,0,0,1) $
                unfoldr reduce (a,b)
in  (x,y)
\end{lstlisting}
All intermediate steps are exposed directly to the compiler, which can optimize more aggressively.

\subsection{Implicit configuration via reflection}
\label{ssec:reflection-package}

This paragraph discusses a problem inherent to interface design in functional programming languages. The reader who is not interested in such details can skip to the next section.

The configuration problem in functional programming is that data that is given on the outer level of a program needs to be accessed in the inner level. Since functional programming style strives to use pure functions---that is, to exclude side effects---this would a priori require one additional configuration parameter in all functions. For example, the two-argument function
\begin{lstlisting}[language=haskell,basicstyle=\small]
f :: a -> b -> c
\end{lstlisting}
would be augmented to
\begin{lstlisting}[language=haskell,basicstyle=\small]
f' :: cfg -> a -> b -> c
\end{lstlisting}
and one would need to carry the argument~\textsf{cfg} through all function calls.

One solution to the dynamic configuration problem proposed in~\cite{implicit_config} is to let the type system assist. One introduces a pair of functions
\begin{lstlisting}[language=haskell,basicstyle=\small]
data Proxy s = Proxy
reify :: a -> (forall (s :: *).
              Reifies s a => Proxy s -> r)
           -> r
reflect :: Reifies s a => Proxy s -> a
\end{lstlisting}
Observe that~\textsf{Proxy} carries no runtime information, since it has one constructor without any parameter. The first argument of \textsf{reify} is a configuration parameter, and its second argument is a function that requires configuration. Inside that function, one can use \textsf{reflect} to recover the configuration parameter from an instance of~\textsf{Proxy}. Prototypical application of this idea would be as follows:
\begin{lstlisting}[language=haskell,basicstyle=\small]
import Data.Reflection
import Data.Proxy
reify 4 $ \(_::Proxy s) ->
          3 + reflect (Proxy :: Proxy s)
\end{lstlisting}
Edward Kmett's library \textsf{reflection} provides a fast implementation of reflection.

\section{HLinear}

We have split HLinear into three packages: algebraic-structures, HFlint, and the main package HLinear. They rely on various Haskell packages authored by others, most prominently the \textsf{vector} package for tuned vector and array manipulation, and Edward Kmett's \textsf{reflection} package. Testing relies on a combination of \textsf{QuickCheck} and \textsf{SmallCheck}, bundled conveniently in the testing framework \textsf{Tasty}. Benchmarking is based on \textsf{Criterion}.

\subsection{algebraic-structures}

The new package algebraic-structures provides classes for algebraic structures ranging from magmas, groups, and actions, to rings, modules, and algebras. For example, magmas are sets together with a binary operator; no further conditions are imposed. The class
\begin{lstlisting}[language=haskell,basicstyle=\small]
class MultiplicativeMagma a where
  (*) :: a -> a -> a
\end{lstlisting}
therefore captures completely the definition of a magma with its binary operator written multiplicatively. At the other extreme, (unital) $R$\nbd algebras can be characterized as sets~$A$ with binary operators $+$ and $\,\cdot\,$, such that
\begin{enumerateroman*}
\item $R$ is commutative,
\item $A$ is a left $R$\nbd algebra,
\item $A$ is a right $R$\nbd algebra,
\item $A$ is an $R$\nbd module.
\end{enumerateroman*}
A type~\textsf{a} representing an algebra with base ring represented by~\textsf{r} is thus described by the class
\begin{lstlisting}[language=haskell,basicstyle=\small]
class ( Commutative r
      , LeftAlgebra r a, RightAlgebra r a
      , Module r a )
  => Algebra r a
\end{lstlisting}

Commutativity is implemented by a similarly-looking class
\begin{lstlisting}[language=haskell,basicstyle=\small]
class MultiplicativeMagma a
  => Commutative a
\end{lstlisting}
However, it includes the axiom that $a_1 \cdot a_2 = a_2 \cdot a_1$ for all instances~$a_1, a_2$ of type~\textsf{a}. There is no general way to establish that such axioms are valid for a given Haskell function. For this reason, the package algebraic-structures provides Tasty-combinators to test that implementations respect relevant mathematical axioms. To invoke only the combinator for commutativity inside a test group, we can write
\begin{lstlisting}[language=haskell,basicstyle=\small]
testGroup "Algebraic properties of a" $
  runTestQSC
  [ isCommutative (Proxy :: Proxy a) ]
\end{lstlisting}

Instances for some built-in Haskell types (Integer, Int, Natural, and Rational) are provided. Two typical instances are given by
\begin{lstlisting}[language=haskell,basicstyle=\small]
mkEuclideanDomainInstanceFromIntegral
  (return []) [t|Integer|]
mkRingInstance (return []) [t|Int|]
\end{lstlisting}
from which the pattern for invoking the corresponding Template Haskell should become clear. As a side remark, note that the type \textsf{Int}, as opposed to \textsf{Integer}, does not represent an integral domain due to overflow.

The package algebraic-structures makes it easier to implement mathematical ideas in greatest possible generality. For example, the normalized PLE~decomposition can be defined for all division rings. And this is exactly the level of generality that HLinear meets. Performance-critical code can be specialized via rewrite rules.

\subsection{HFlint}

HFlint is a wrapper around some parts of Flint. Specifically, it wraps integers~$\ZZ$, rationals~$\QQ$, polynomial algebras~$\ZZ[x]$ and~$\QQ[x]$, finite fields~$\bbF_p$ for primes~$p$, and number fields (via Antic~\cite{antic}). Given the current capabilities of Flint, it is possible to extend this to $p$-adics ($\QQ_p$, $\ZZ_p$, and their finite extensions), to all finite fields $\bbF_q$ for prime powers~$q$, and to the real and complex numbers~$\RR$ and~$\CC$ (via Arb~\cite{arb}). In light of the layout of HFlint, this extension would be feasible with rather little effort.

In Section~\ref{ssec:background-implementations}, we have discussed low-level and high-level optimizations of linear algebra. While papers like~\cite{mainland-stream-fusion,sheeran-fpga-layout} suggest that the performance barrier between C and Haskell for elementary operations is low or even non-existent, state-of-the-art implementations of, say, integer arithmetic (GMP, MPIR) are very hard to beat. This shall not be the topic of this paper. Consequently, we rely on Flint for fundamental
arithmetic. Note that we purposely wrap~$\ZZ$, which is also represented by \textsf{Integer}. However, \textsf{Rational} (built on top of \textsf{Integer}) cannot compete with the corresponding Flint implementation.

Flint objects are either synonyms for an elementary data type (\textsf{int}, \textsf{long}, etc.) or pointers to C structures. Functions for Flint objects may accept a context object storing information about the type rather than an individual object. For example, rationals are usually accessed via
\begin{lstlisting}[language=c,basicstyle=\small]
typedef fmpq fmpq_t[1];
\end{lstlisting}
and the signature of a typical function is
\begin{lstlisting}[language=c,basicstyle=\small]
void fmpq_add(fmpq_t res,
  const fmpq_t op1, const fmpq_t op2);
\end{lstlisting}
with self-evident meaning of the arguments. There are no context objects attached to rationals.

Finite fields~$\bbF_p$ do have a context associated to them that keeps track of the prime~$p$. Elements of finite fields of small moduli are encoded by means of an elementary data type \textsf{mp\_limb\_t}, which on most common architectures resolves as \textsf{unsigned long}. A typical signature for such finite fields is
\begin{lstlisting}[language=c,basicstyle=\small]
mp_limb_t nmod_add(
    mp_limb_t a, mp_limb_t b, nmod_t mod)
\end{lstlisting}
The first two arguments are the summands and the third one is a pointer to a context.

The disciplined interface style of Flint allows for systematic wrapping. As mentioned above there is a context attached to finite fields. Clearly it is desirable to disallow, say, addition of elements of $\bbF_p$ and $\bbF_{p'}$ for $p \ne p'$. A context parameter on the left hand side of the data type declaration takes care of this. Note that the context does not appear on the right.
\begin{lstlisting}[language=haskell,basicstyle=\small]
type FlintLimb = CULong
newtype NMod ctxProxy =
  NMod {unNMod :: FlintLimb}
\end{lstlisting}

To operate with elements of finite fields, Flint requires the context reference~\textsf{nmod\_t}. The implementation of addition with type signature
\begin{lstlisting}[language=haskell,basicstyle=\small]
(+) :: NMod ctx -> NMod ctx -> NMod ctx
\end{lstlisting}
can therefore be viewed as a dynamic configuration problem. We make \textsf{NMod} an instance of the class~\textsf{FlintPrim}, which contains the function
\begin{lstlisting}[language=haskell,basicstyle=\small]
withFlintPrimCtx
  :: ReifiesFlintContext ctx ctxProxy
  => NMod ctxProxy
  -> (    CFlintPrim NMod
       -> Ptr (CFlintCtx ctx) -> IO b)
  -> IO b
\end{lstlisting}
The condition~\textsf{ReifiesFlintContext} is related to the reflection library and is discussed in the next subsection. The second argument corresponds to a wrapped C function. Concretely, addition could be implemented as
\begin{lstlisting}[language=haskell,basicstyle=\small]
(+) a b =
  withFlintPrimCtx a $ \ac _      ->
  withFlintPrimCtx b $ \bc ctxptr ->
  nmod_add ac bc ctxptr
\end{lstlisting}
The inner variables \textsf{ac} and \textsf{bc} arise from the data represented by~\textsf{a} and~\textsf{b}, but the context originates from their type via reflection. In particular, it is semantically correct to ignore the context pointer provided by the first call to \textsf{withFlintPrimCtx}, since~\textsf{a} and~\textsf{b} have the same type, and therefore they have the same context.

To create and employ the context of a finite field one uses \textsf{withNModContext} or \textsf{withNModContextM}. The first argument to them is the modulus of the finite field. The type signature of \textsf{withNModContext} includes the condition~\textsf{NFData b} for the type~\textsf{b} of the result.
\begin{lstlisting}[language=haskell,basicstyle=\small]
withNModContext
 :: NFData b
 => FlintLimb
 -> (    forall ctxProxy .
         ReifiesNModContext ctxProxy
      => Proxy ctxProxy -> b)
 -> b

withNModContextM
 :: ( MonadIO m, MonadMask m )
 => FlintLimb
 -> (    forall ctxProxy .
         ReifiesNModContext ctxProxy
      => Proxy ctxProxy -> m b)
 -> m b
\end{lstlisting}

\subsubsection*{Reflection and dynamic types}
\label{ssec:refection-and-dynamic-types}

The goal of this section is to explain the condition~\textsf{NFData b} in the type signature of \textsf{withNModContext}, which might appear unnecessary. Context objects are generally represented by a pointer to a C structure. Most commonly, one would use a \textsf{ForeignPtr}, whose finalizer frees the C structure. A typical computation in~$\bbF_7$ might look as follows.
\begin{lstlisting}[language=haskell,basicstyle=\small]
unsafePerformIO $ do
  ctx <- newFlintContext $ NModCtxData 7
  withFlintContext ctx $
    unNMod $ (NMod 3) + (NMod 4)
\end{lstlisting}
Invoking in this way an implementation based on foreign pointers and Kmett's \textsf{reflection} library can and will produce segmentation faults. The use of \textsf{unsafePerformIO} in the C-bindings is at the root of this problem.

The concept of reflection is mathematically sound, but is incompatible with finalizers when performed unsafely. A priori, the inner function in the second line does not contain any reference to the C context instance. This might make the finalizer free it before an actual reference is created by
\begin{lstlisting}[language=haskell,basicstyle=\small]
reflect (Proxy :: Proxy ctx)
\end{lstlisting}
The point is that the latter call does create runtime data out of type information, which the finalizer of a \textsf{ForeignPtr} cannot keep track of.

Dynamic configuration via reflection is fast, since it allows us to move context information away from the element information. Instances of a hypothetical data type
\begin{lstlisting}[language=haskell,basicstyle=\small]
data NMod' = NMod' NModContext FlintLimb
\end{lstlisting}
will not only double the memory footprint, but also prevent some optimizations for the elementary data type \textsf{FlintLimb}.

Dynamic configuration is also convenient, since it transparently prevents accidental combination of elements of different fields. It thus seems worth to introduce an \textsf{NFData b} condition to maintain it. Context objects will be implemented by plain \textsf{Ptr} instances, and memory allocation should be taken care of manually. The implementation of~\textsf{withNModContext} in HFlint is
\begin{lstlisting}[language=haskell,basicstyle=\small]
withNModContext n f = unsafePerformIO $ do
  ctx <- newFlintContext $ NModCtxData n
  let h = force $ withFlintContext ctx f
  seq h $ freeFlintContext ctx
  return h
\end{lstlisting}

\subsection{HLinear}

HLinear implements arithmetic with matrices over arbitrary rings, and PLE decomposition of matrices over division rings. It makes heavy use of the \textsf{vector} library, which optimizes function applications and stores elementary data types efficiently.

Matrices keep track (as runtime data) of their dimensions, and their entries as vectors of rows in the same way that Flint does. The corresponding data declaration is
\begin{lstlisting}[language=haskell,basicstyle=\small]
data Matrix a =
  Matrix { nmbRows :: Int
         , nmbCols :: Int
         , rows    :: Vector (Vector a)
         }
\end{lstlisting}
Matrix arithmetic is performed in a straightforward way. In particular, we do not employ Strassen multiplication\footnote{Since classical Gaussian elimination does not use matrix multiplication, this choice is of no consequence for the focus of the paper.}.

\subsubsection*{Gaussian elimination}
Most of the implementation is devoted to the PLE decomposition based on Gaussian elimination.
We illustrate it by returning to the example in Section~\ref{ssec:ple-decomposition}.
If the first column of $M$ is non-zero (which is the case in Section~\ref{ssec:ple-decomposition}), we can write~$M$ as a product
\begin{gather}
\label{eq:split-off-hook}
  M = P L (E + M')
\end{gather}
such that $P$ is a permutation matrix and
\begin{gather*}
  L=\begin{psmatrix}
  \ast' & 0 & 0 & 0 \\
  \ast  & 1 & 0 & 0 \\
  \ast  & 0 & 1 & 0 \\
  \ast  & 0 & 0 & 1 \\
  \end{psmatrix}
  \tx{,}
  \enskip
  E=\begin{psmatrix}
  1 & \ast & \ast & \ast & \ast & \ast \\
  0 & 0    & 0    & 0    & 0    & 0    \\
  0 & 0    & 0    & 0    & 0    & 0    \\
  0 & 0    & 0    & 0    & 0    & 0    \\
  \end{psmatrix}
  \tx{,}
  \enskip
  M'=\begin{psmatrix}
      0 & 0 & 0 & 0 & 0 & 0\\
      0\\
      0 & \multicolumn{5}{c}{$M'_0$}\\
      0\\
    \end{psmatrix}
  \tx{.}
\end{gather*}
Suppose we have a PLE~decomposition of the smaller matrix $M'_0=P'_0 L'_0 E'_0$.
This naturally gives rise to a decomposition $M'=P'L'E'$ by
\begin{enumerateroman*}
\item letting $P'$ be the permutation matrix that, when acting from the left, fixes the top row and acts on the remaining rows like $P'_0$;
\item setting $L'$ and $E'$ to
\end{enumerateroman*}
\begin{gather*}
  L'=\begin{psmatrix}
      1 & 0 & 0 & 0\\
      0\\
      0 & \multicolumn{3}{c}{$L'_0$}\\
      0\\
    \end{psmatrix}
  \tx{,}
  \quad
  E'=\begin{psmatrix}
      0 & 0 & 0 & 0 & 0 & 0\\
      0\\
      0 & \multicolumn{5}{c}{$E'_0$}\\
      0\\
    \end{psmatrix}
  \tx{.}
\end{gather*}

We thus obtain a decomposition
\begin{gather*}
  M = P L (E + P' L' E')
\tx{.}
\end{gather*}
Notice that we have $E = P' L' E$, since both $P'$ and $L'$ fix the top row when acting from the left. In particular, we obtain a PLE decomposition
\begin{multline}
\label{eq:combine-hooks}
  M
=
  P L (E + P' L' E')
=
  P L (P' L' E + P' L' E')
\\
=
  P P' \big( P^{\prime\, -1} L P' \big) L' (E + E')
\tx{.}
\end{multline}
It is important to observe that $P^{\prime\, -1} L P'$ is a lower triangular matrix.

The heart of HLinear is about effectively modeling the computation that we just described. We call a triple of matrices $(P, L, E)$ as in~\eqref{eq:split-off-hook} a PLE~hook.
\begin{definition}
A PLE~hook of size~$n$ is a triple $(P,L,E)$ of a permutation $P$ on $n$ rows, a lower triangular matrix~$L$ of size $n \times n$, and a matrix~$E$ with $n$ rows that is a row sum of a zero matrix and a matrix in echelon form.

We say that the PLE~triple $(P,L,E)$ has rank~$r$ and corank~$r'$ if
\begin{enumerateroman*}
\item $P - I_n$ is supported on rows and colums of indices $n - r - r' < j$;
\item $L - I_n$ is supported on columns of indices  $n - r - r' < j \le n - r'$; and
\item $E$ is supported on rows of indices  $n - r - r' < i \le n - r'$.
\end{enumerateroman*}
\end{definition}
\begin{example}
The triple of matrices
\begin{gather*}
  \begin{psmatrix}
   1 & 0 & 0 & 0 \\
   0 & 0 & 0 & 1 \\
   0 & 1 & 0 & 0 \\
   0 & 0 & 1 & 0
  \end{psmatrix}
\tx{,}\quad
  \begin{psmatrix}
   1 & 0 & 0 & 0 \\
   0 & \frac{1}{2} & 0 & 0 \\
   0 & -3 & 4 & 0 \\
   0 & \frac{7}{3} & -17 & 1
  \end{psmatrix}
\tx{,}\quad\tx{and}\quad
  \begin{psmatrix}
  0 & 0 & 0 & 0 & 0 & 0 \\
  0 & 0 & 0 & 1 & -3 & \frac{1}{13} \\
  0 & 0 & 0 & 0 & 1 & 0 \\
  0 & 0 & 0 & 0 & 0 & 0 \\
  \end{psmatrix}
\end{gather*}
is a PLE~hook of size~$4$, rank~$2$, and corank~$1$.
\end{example}

The corresponding data type in Haskell is
\begin{lstlisting}[language=haskell,basicstyle=\small]
data PLEHook a b =
  PLEHook
    RPermute
    (LeftTransformation a)
    (EchelonForm b)
\end{lstlisting}
Data types for permutations, left transformations, and echelon forms will be explained below.

Splitting off the first non-zero column and regarding the remaining bottom right matrix as a matrix of smaller size yields a list of PLE~hooks when iterating. In Haskell, we can obtain this list as
\begin{lstlisting}[language=haskell,basicstyle=\small]
unfoldr splitOffHook
\end{lstlisting}
based on the function
\begin{lstlisting}[language=haskell,basicstyle=\small]
splitOffHook
 :: ( DivisionRing a
    , DecidableZero a, DecidableUnit a )
 => Matrix a -> Maybe (PLEHook a,Matrix a)
\end{lstlisting}

Equation~\eqref{eq:combine-hooks} in the above discussion of PLE~hooks yields a partially-defined associative product on PLE~hooks. Given PLE~hooks $(P_1,L_1,E_1)$ and $(P_2,L_2,E_2)$ of ranks $r_1$, $r_2$ and coranks~$r'_1$ and $r'_2$, we set
\begin{multline*}
  (P_1,L_1,E_1) \cdot (P_2,L_2,E_2)
\\
=
  \big( P_1 P_2,\, P_2^{-1} L_1 P_2 L_2,\, E_1 + E_2 \big)
\end{multline*}
if $r'_1 \ge r_2 + r'_2$. One verifies that this condition is satisfied for all sequences of PLE~hooks that arise from unfolding a matrix with \textsf{splitOffHook}. As a result, we can formulate PLE~decomposition of a matrix~\textsf{m} as
\begin{lstlisting}[language=haskell,basicstyle=\small]
foldl (*) (firstHook nrs ncs) $
  unfoldr splitOffHook m
\end{lstlisting}

\subsubsection*{PLE hooks}
\label{ssec:ple-hooks}

PLE~hooks consist of three elements, $P$, $L$, and $E$. The implementation is designed for general division rings, with no particular optimizations for rational numbers.

For permutations, we rely on the library \textsf{permutation}, which internally makes use of \textsf{IntArray}. Their action on \textsf{Vector} is implemented directly by invoking functionality of the \textsf{vector} library. Since permutations are only a small part of the computation, we do not discuss them in more detail. Efficient data types for $L$ and $E$, on the other hand, are crucial to good overall performance.

Matrices $L$ are encoded as left transformations, whose declaration is
\begin{lstlisting}[language=haskell,basicstyle=\small]
data LeftTransformation a =
  LeftTransformation
    { nmbRows :: Int
    , columns :: Vector
          (LeftTransformationColumn a)
    }
data LeftTransformationColumn a =
  LeftTransformationColumn
    { offset :: Int
    , headUnit :: Unit a
    , tail :: Vector a
    }
\end{lstlisting}
The first parameter of \textsf{LeftTransformation} refers to the number of rows of~$L$, which is the same as its number of columns. The second parameter is a list of columns, not necessarily exhausting all. For example, a left transformation with $4$ rows and only two columns listed will be of the form
\begin{gather*}
  \begin{psmatrix}
  \ast' & 0     & 0 & 0 \\
  \ast  & \ast' & 0 & 0 \\
  \ast  & \ast  & 1 & 0 \\
  \ast  & \ast  & 0 & 1 \\
  \end{psmatrix}
\tx{.}
\end{gather*}
Columns of left transformations keep track of their column index~$j$, which is referred to as~\textsf{offset} (from the top). The \textsf{headUnit} of a column is its $j$\thdash\ element. The newtype wrapper \textsf{Unit} ensures that it is not zero, which over division rings is equivalent to being invertible. All remaining entries of a left transformation column are stored in a vector~\textsf{tail}.

Implementing left transformations with offsets ensures that they are stored as compactly as possible. The separate saving of the column index, which would a priori be deducible from the container~\textsf{columns}, makes some operations more localizable.

Echelon forms are stored in a way that is similar to left transformations.
\begin{lstlisting}[language=haskell,basicstyle=\small]
data EchelonForm a =
  EchelonForm
    { nmbRows :: Int
    , nmbCols :: Int
    , rows :: Vector (EchelonFormRow a)
    }
data EchelonFormRow a =
  EchelonFormRow
    { offset :: Int
    , row :: Vector a
    }
\end{lstlisting}
We need to keep track of both the number of rows and columns of $E$. A row of vectors has an offset as left transformation columns do.

We conclude with an observation. From experience with HLinear it seems that its performance can profit from intermediate objects that are defined at mathematical level of rigor. If true, the roots of this observation might be the compiler's ability to rearrange intermediate steps more effectively, i.e.\ to optimize more aggressively. It can definitely not be related to rewrite rules, which are not included in the current version of algebraic-structures.

\subsection{Reduction of echelon forms}

We treat the reduction of echelon forms briefly, appealing to the analogy with Section~\ref{ssec:ple-hooks}. The row reduced echelon form of a matrix~$M$ can be obtained from the echelon form~$E$ in a PLE~decomposition by applying upper triangular matrices from the right. In analogy to the implementation of the PLE~decomposition we implement a PLUE~decomposition in which~$P$ is a permutation, $L$ is a left transformation as before, $U$ is an upper triangular matrix with ones on its diagonal, and $E$ is a row reduced echelon form. A PLUE~decomposition $M = PLUE'$ can be computed from a PLE~decomposition $M = PLE$ by reducing $E = U E'$. This reduction is based on a fold-unfold algorithm as before. The foundational data type is
\begin{lstlisting}[language=haskell,basicstyle=\small]
data ERHook a =
  ERHook
    (EchelonTransformation a)
    (Matrix a)
    (EchelonForm a)
\end{lstlisting}
A reduction step for an echelon form $E$ consists of splitting off the rightmost pivot element
\begin{gather*}
  E
=
  \begin{psmatrix}
  E_\rml & M_\rmr \\
  0 & E_\rmr
  \end{psmatrix}
\end{gather*}
and computing a transformation $U$ that reduces the first column of the second block-column:
\begin{gather*}
  U
  \begin{psmatrix}
  M_\rmr \\
  E_\rmr
  \end{psmatrix}
=
  \begin{psmatrix}
  M'_\rmr \\
  E_\rmr
  \end{psmatrix}
\tx{.}
\end{gather*}
Here, the first column of $M'_\rmr$ is zero. The resulting ER~hook then consists of $(U, M'_\rmr, E_\rmr)$ and the remaining echelon form is $E_\rml$. The product of ER~hooks is given by
\begin{gather*}
  (U, M, E)
  \cdot
  (U', M', E')
=
  \Bigg(
  \begin{psmatrix}
  U U'_\rmt \\ U'_\rmb
  \end{psmatrix}
  ,\;
  \begin{psmatrix}
  M & U M'_\rmt
  \end{psmatrix}
  ,\;
  \begin{psmatrix}
  E & M'_\rmb  \\
  0 & E'
  \end{psmatrix}
  \Bigg)
\tx{,}
\end{gather*}
where $M' = \rT \begin{psmatrix} M'_\rmt & M'_\rmb \end{psmatrix}$ and $U' = \rT \begin{psmatrix} U'_\rmt & U'_\rmb \end{psmatrix}$ are vertical decompositions of $M'$ and $U'$ that are compatible with the size of~$M$.

% \subsection{Correctness}

% \remarkil{insert}

\section{Usage}

We illustrate usage of HLinear via the computation of one example.
\begin{lstlisting}[language=haskell,basicstyle=\small]
import HFlint.FMPQ
import HLinear.Matrix as M
import HLinear.NormalForm

let m = M.fromLists
  [[ 84 , 168 , 588 ,-252 , 336 , 49   ]
  ,[ 672,1344 ,4704 ,-1992,4722 ,2552  ]
  ,[-504,-1008,-3528,2100 ,-1575,-4998 ]
  ,[ 168, 336 ,1176 ,-168 ,1428 ,-2002]]
  :: Matrix FMPQ
let [p,l,e] = toMatrices $
  ple m :: [Matrix FMPQ]
\end{lstlisting}

Reformatting the output slightly, this yields
\begin{gather*}
  \begin{psmatrix}
   1 & 0 & 0 & 0 \\
   0 & 1 & 0 & 0 \\
   0 & 0 & 1 & 0 \\
   0 & 0 & 0 & 1
  \end{psmatrix}
\tx{,}\;
  \begin{psmatrix}
   84  &  0  &   0    & 0 \\
  672  &  24 &   0    & 0 \\
  -504 & 588 & -49392 & 0 \\
  168  & 336 & -27720 & 1
  \end{psmatrix}
\tx{,}\;
  \begin{psmatrix}
   1 & 2 & 7 &-3 & 4 & 72 \\
   0 & 0 & 0 & 1 &\frac{339}{4} & 90 \\
   0 & 0 & 0 & 0 & 1 & \frac{7}{6} \\
   0 & 0 & 0 & 0 & 0 & 0
  \end{psmatrix}
\tx{.}
\end{gather*}

Behind the scenes, \textsf{NF.ple} invokes the fold-unfold implementation of PLE decomposition. It returns a PLE~hook object, which can be converted to the list of matrices $P,L,E$ via \textsf{toMatrices}.

The row reduced echelon form of the matrix~$m$ above can be obtained by
\begin{lstlisting}[language=haskell,basicstyle=\small]
let [p,l,u,e] = toMatrices $
  rref m :: [Matrix FMPQ]
\end{lstlisting}

\section{Performance}

%% PERFORMANCE AS READ BY JULIA SCRIPT
%
%  randMat_nrs10_ncs10_snum10_nden5_sden2
%  [0.20967, 0.0479377]
%  [0.313779, 0.0317065]
%
%  randMat_nrs10_ncs20_snum10_nden5_sden2
%  [0.20894, 0.130442]
%  [2.69408, 1.15349]
%
%  randMat_nrs10_ncs30_snum10_nden5_sden3
%  [0.355645, 0.410266]
%  [9.35648, 7.77253]
%
%  randMat_nrs10_ncs40_snum10_nden5_sden4
%  [0.542425, 0.906161]
%  [18.6526, 20.5843]
%
%  randMat_nrs10_ncs10_snum50_nden5_sden5
%  [1.05333, 0.214856]
%  [1.5632, 0.157513]
%
%  randMat_nrs10_ncs20_snum50_nden5_sden5
%  [1.0794, 0.530794]
%  [14.5898, 5.87683]
%
%  randMat_nrs10_ncs30_snum50_nden5_sden5
%  [1.0576, 0.879937]
%  [24.1065, 15.1527]
%
%  randMat_nrs20_ncs20_snum50_nden5_sden5
%  [49.5401, 10.3247]
%  [103.434, 10.111]
%
%  randPLE_nrs10_ncs10_snum10_nden5_sden2
%  [0.0151787, 0.0202448]
%  [0.0137755, 0.0183578]
%
%  randPLE_nrs10_ncs20_snum10_nden5_sden2
%  [0.0510828, 0.240302]
%  [0.0539852, 0.299181]
%
%  randPLE_nrs10_ncs30_snum10_nden5_sden3
%  [0.0970813, 0.842287]
%  [0.085508, 1.1337]
%
%  randPLE_nrs10_ncs40_snum10_nden5_sden4
%  [0.380611, 4.53674]
%  [0.360511, 6.09929]
%
%  randPLE_nrs10_ncs10_snum50_nden5_sden5
%  [0.129645, 0.0840513]
%  [0.170735, 0.0884088]
%
%  randPLE_nrs10_ncs20_snum50_nden5_sden5
%  [0.35502, 1.2109]
%  [0.318277, 1.47072]
%
%  randPLE_nrs10_ncs30_snum50_nden5_sden5
%  [0.526052, 2.83387]
%  [0.512875, 4.35638]
%
%  randPLE_nrs20_ncs20_snum50_nden5_sden5
%  [0.448953, 1.16507]
%  [0.616029, 1.41178]
%
%  randPLE_nrs60_ncs60_snum1_nden1_sden1
%  [0.190447, 1.98559]
%  [0.162083, 2.18579]
%
%  randPLE_nrs100_ncs100_snum1_nden1_sden1
%  [1.00079, 17.3218]
%  [0.750342, 14.6112]

We compared the performance of HLinear to that of Flint/Nemo\footnote{Flint is wrapped thinly by its authors in Nemo.} via a suite of benchmarks (all run on one core of a 3.00GHz Intel Xeon E5-2683 processor with 12~GB RAM shared by 56 cores).

\subsection*{Random matrices}

Our benchmarks use
\begin{enumeratearabic*}
\item square matrices with random entries, and
\item the product of a random permutation, and a left transformation and an echelon form as in Section~\ref{ssec:ple-decomposition} with random entries.
\end{enumeratearabic*}
We refer to the latter as random PLE matrices to distinguish them. To accommodate our focus on matrices with bounded denominators, we generate matrix entries whose denominators are products $d_1 \cdots d_n$ for random numbers~$d_i$. We use the following parameters:
\begin{description}
\setlength\itemsep{-.3em}
  \item[nrs, ncs] matrix size;
  \item[snum] upper bound on the size of the numerators of the entries (in words, i.e.\ multiples of 8~bytes);
  \item[nden] upper bound on the number of factors used to generate denominators of the entries;
  \item[sden] upper bound on the size of the factors used to generate denominators of the entries (in words).
\end{description}
Notice that the PLE decomposition of random matrices typically contains large denominators, while this is not expected for random PLE matrices.

Both Flint and HLinear are run on the same random matrices. Tables~\ref{tab:performance:mat-classical} and~\ref{tab:performance:ple-classical} show results for various parameters at a time. For each combination of parameters we generate several matrices and average their benchmarks. 

\begin{table}[htbp]
\caption{Classical Gaussian elimination on random matrices}
\label{tab:performance:mat-classical}
  \setlength{\tabcolsep}{0.3em}
  \begin{tabular}{ccccrr}
  \toprule
  $\mathrm{nrs}$ & $\mathrm{ncs}$ & $\mathrm{snum}$ & $\mathrm{nden} \times \mathrm{sden}$ &
    \multicolumn{2}{c}{CPU time in ms} \\
   & & & & HLinear & Flint \\
  \midrule
  $10$ & $10$ & $10$ & $5 \times 2 = 10$ &
    \textbf{\numprint{210}} & \numprint{314} \\
  $10$ & $20$ & $10$ & $5 \times 2 = 10$ &
    \textbf{\numprint{209}} &  \numprint{2694} \\
  $10$ & $30$ & $10$ & $5 \times 3 = 15$ &
    \textbf{\numprint{356}} & \numprint{9356} \\
  $10$ & $40$ & $10$ & $5 \times 4 = 20$ &
    \textbf{\numprint{542}} & \numprint{18653} \\
  $10$ & $10$ & $50$ & $5 \times 5 = 25$ &
    \textbf{\numprint{1053}} & \numprint{1563} \\
  $10$ & $20$ & $50$ & $5 \times 5 = 25$ &
    \textbf{\numprint{1079}} & \numprint{14590} \\
  $10$ & $30$ & $50$ & $5 \times 5 = 25$ &
    \textbf{\numprint{1058}} & \numprint{24107} \\
  $20$ & $20$ & $50$ & $5 \times 5 = 25$ &
    \textbf{\numprint{49540}} & \numprint{103434} \\
  \bottomrule
\end{tabular}
\end{table}

\begin{table}[htbp]
\caption{Classical Gaussian elimination on random PLE matrices}
\label{tab:performance:ple-classical}
  \setlength{\tabcolsep}{0.3em}
  \begin{tabular}{ccccrr}
  \toprule
  $\mathrm{nrs}$ & $\mathrm{ncs}$ & $\mathrm{snum}$ & $\mathrm{nden} \times \mathrm{sden}$ &
    \multicolumn{2}{c}{CPU time in ms} \\
   & & & & HLinear & Flint \\
  \midrule
  $10$ & $10$ & $10$ & $5 \times 2 = 10$ &
    \numprint{15} & \textbf{\numprint{14}} \\
  $10$ & $20$ & $10$ & $5 \times 2 = 10$ &
    \textbf{\numprint{51}} & \numprint{54} \\
  $10$ & $30$ & $10$ & $5 \times 3 = 15$ &
    \numprint{97} & \textbf{\numprint{86}} \\
  $10$ & $40$ & $10$ & $5 \times 4 = 20$ &
    \numprint{381} & \textbf{\numprint{361}} \\
  $10$ & $10$ & $50$ & $5 \times 5 = 25$ &
    \textbf{\numprint{130}} & \numprint{171} \\
  $10$ & $20$ & $50$ & $5 \times 5 = 25$ &
    \numprint{355} & \textbf{\numprint{318}} \\
  $10$ & $30$ & $50$ & $5 \times 5 = 25$ &
    \numprint{526} & \textbf{\numprint{513}} \\
  $20$ & $20$ & $50$ & $5 \times 5 = 25$ &
    \textbf{\numprint{449}} & \numprint{616} \\
  $60$ & $60$ & $1$ & $1 \times 1 = 1$ &
    \numprint{190} & \textbf{\numprint{162}} \\
  $100$ & $100$ & $1$ & $1 \times 1 = 1$ &
    \numprint{1001} & \textbf{\numprint{750}} \\
  \bottomrule
\end{tabular}
\end{table}

\subsection*{Classical and fraction-free Gaussian elimination}

Flint implements three variants of reduced row echelon form (rref), based on classical Gaussian elimination, on a fraction-free variant~\cite{bareiss-1968}, and on multi-modular arithmetic. One of the referees pointed out that, for several parameter sets in these benchmarks, classical Gaussian elimination is inferior to fraction-free Gaussian elimination; and therefore that we should be comparing to the fraction-free variant. Indeed, the number of rows of test matrices is seriously limited when restricting to the classical algorithm. In reaction to this critique and in order to complete the discussion, we have provided a coarse implementation of fraction-free Gaussian elimination along the lines of the one in Flint. Tables~\ref{tab:performance:mat-fractionfree} and~\ref{tab:performance:ple-fractionfree} show benchmark results for various parameters.

\begin{table}[htbp]
\caption{Fraction-free Gaussian elimination on random matrices}
\label{tab:performance:mat-fractionfree}
  \setlength{\tabcolsep}{0.3em}
  \begin{tabular}{ccccrr}
  \toprule
  $\mathrm{nrs}$ & $\mathrm{ncs}$ & $\mathrm{snum}$ & $\mathrm{nden} \times \mathrm{sden}$ &
    \multicolumn{2}{c}{CPU time in ms} \\
   & & & & HLinear & Flint \\
  \midrule
  $10$ & $10$ & $10$ & $5 \times 2 = 10$ &
    \numprint{48} & \textbf{\numprint{32}} \\
  $10$ & $20$ & $10$ & $5 \times 2 = 10$ &
    \textbf{\numprint{130}} & \numprint{1153} \\
  $10$ & $30$ & $10$ & $5 \times 3 = 15$ &
    \textbf{\numprint{410}} & \numprint{7773} \\
  $10$ & $40$ & $10$ & $5 \times 4 = 20$ &
    \textbf{\numprint{906}} & \numprint{20584} \\
  $10$ & $10$ & $50$ & $5 \times 5 = 25$ &
    \numprint{215} & \textbf{\numprint{158}} \\
  $10$ & $20$ & $50$ & $5 \times 5 = 25$ &
    \textbf{\numprint{531}} & \numprint{5877} \\
  $10$ & $30$ & $50$ & $5 \times 5 = 25$ &
    \textbf{\numprint{880}} & \numprint{15153} \\
  $20$ & $20$ & $50$ & $5 \times 5 = 25$ &
    \numprint{10325} & \textbf{\numprint{10111}} \\
  \bottomrule
\end{tabular}
\end{table}

\begin{table}[htbp]
\caption{Fraction-free Gaussian elimination on random PLE matrices}
\label{tab:performance:ple-fractionfree}
  \setlength{\tabcolsep}{0.3em}
  \begin{tabular}{ccccrr}
  \toprule
  $\mathrm{nrs}$ & $\mathrm{ncs}$ & $\mathrm{snum}$ & $\mathrm{nden} \times \mathrm{sden}$ &
    \multicolumn{2}{c}{CPU time in ms} \\
   & & & & HLinear & Flint \\
  \midrule
  $10$ & $10$ & $10$ & $5 \times 2 = 10$ &
    \numprint{20} & \textbf{\numprint{18}} \\
  $10$ & $20$ & $10$ & $5 \times 2 = 10$ &
    \textbf{\numprint{24}} & \numprint{30} \\
  $10$ & $30$ & $10$ & $5 \times 3 = 15$ &
    \textbf{\numprint{842}} & \numprint{1134} \\
  $10$ & $40$ & $10$ & $5 \times 4 = 20$ &
    \textbf{\numprint{4537}} & \numprint{6099} \\
  $10$ & $10$ & $50$ & $5 \times 5 = 25$ &
    \textbf{\numprint{84}} & \numprint{88} \\
  $10$ & $20$ & $50$ & $5 \times 5 = 25$ &
    \textbf{\numprint{1211}} & \numprint{1471} \\
  $10$ & $30$ & $50$ & $5 \times 5 = 25$ &
    \textbf{\numprint{2834}} & \numprint{4356} \\
  $20$ & $20$ & $50$ & $5 \times 5 = 25$ &
    \textbf{\numprint{1165}} & \numprint{1412} \\
  $60$ & $60$ & $1$ & $1 \times 1 = 1$ &
    \textbf{\numprint{1986}} & \numprint{2186} \\
  $100$ & $100$ & $1$ & $1 \times 1 = 1$ &
    \numprint{17322} & \textbf{\numprint{14611}} \\
  \bottomrule
\end{tabular}
\end{table}

\subsection{Thread Caching Malloc}

As opposed to Flint, HLinear does not use in-place arithmetic. Since HFlint does not expose functionality for fusion, we expect that memory allocation and de\-allocation happen significantly more often in HLinear than in Flint. However, profiling runs reveal that only 2\% of the runtime is spent on memory management. To verify that memory management does not play a major role in HLinear, we ran all the previous benchmarks with preloaded \textsf{libtcmalloc}. Supporting our claim, the results in Table~\ref{tab:performance:mat-tcmalloc} and~\ref{tab:performance:ple-tcmalloc} show slightly improved performance, but no major change of runtimes.

We have run Flint benchmarks as part of Nemo, which is a Julia wrapper written by Flint maintainer William Hart. Nemo sets the Flint memory management functions to those of Julia, in order to improve performance compared to system malloc. For this reason we do not benchmark Nemo with thread caching malloc.

\begin{table}[htbp]
\caption{HLinear Gaussian elimination on random matrices using Thread Caching Malloc}
\label{tab:performance:mat-tcmalloc}
  \setlength{\tabcolsep}{0.3em}
  \begin{tabular}{ccccrr}
  \toprule
  $\mathrm{nrs}$ & $\mathrm{ncs}$ & $\mathrm{snum}$ & $\mathrm{nden} \times \mathrm{sden}$ &
    \multicolumn{2}{c}{CPU time in ms} \\
   & & & & Class. & Frac.\ Free \\
  \midrule
  $10$ & $10$ & $10$ & $5 \times 2 = 10$ &
    \numprint{197} & \numprint{46} \\
  $10$ & $20$ & $10$ & $5 \times 2 = 10$ &
    \numprint{195} &  \numprint{123} \\
  $10$ & $30$ & $10$ & $5 \times 3 = 15$ &
    \numprint{337} & \numprint{399} \\
  $10$ & $40$ & $10$ & $5 \times 4 = 20$ &
    \numprint{499} & \numprint{845} \\
  $10$ & $10$ & $50$ & $5 \times 5 = 25$ &
    \numprint{981} & \numprint{207} \\
  $10$ & $20$ & $50$ & $5 \times 5 = 25$ &
    \numprint{1002} & \numprint{502} \\
  $10$ & $30$ & $50$ & $5 \times 5 = 25$ &
    \numprint{1010} & \numprint{849} \\
  $20$ & $20$ & $50$ & $5 \times 5 = 25$ &
    \numprint{47355} & \numprint{9795} \\
  \bottomrule
\end{tabular}
\end{table}

\begin{table}[htbp]
\caption{HLinear Gaussian elimination on random PLE matrices using Thread Caching Malloc}
\label{tab:performance:ple-tcmalloc}
  \setlength{\tabcolsep}{0.3em}
  \begin{tabular}{ccccrr}
  \toprule
  $\mathrm{nrs}$ & $\mathrm{ncs}$ & $\mathrm{snum}$ & $\mathrm{nden} \times \mathrm{sden}$ &
    \multicolumn{2}{c}{CPU time in ms} \\
   & & & & Class. & Frac.\ Free \\
  \midrule
  $10$ & $10$ & $10$ & $5 \times 2 = 10$ &
    \numprint{14} & \numprint{20} \\
  $10$ & $20$ & $10$ & $5 \times 2 = 10$ &
    \numprint{51} & \numprint{24} \\
  $10$ & $30$ & $10$ & $5 \times 3 = 15$ &
    \numprint{90} & \numprint{919} \\
  $10$ & $40$ & $10$ & $5 \times 4 = 20$ &
    \numprint{362} & \numprint{4238} \\
  $10$ & $10$ & $50$ & $5 \times 5 = 25$ &
    \numprint{137} & \numprint{82} \\
  $10$ & $20$ & $50$ & $5 \times 5 = 25$ &
    \numprint{343} & \numprint{1096} \\
  $10$ & $30$ & $50$ & $5 \times 5 = 25$ &
    \numprint{489} & \numprint{2788} \\
  $20$ & $20$ & $50$ & $5 \times 5 = 25$ &
    \numprint{435} & \numprint{1155} \\
  $60$ & $60$ & $1$ & $1 \times 1 = 1$ &
    \numprint{185} & \numprint{2061} \\
  $100$ & $100$ & $1$ & $1 \times 1 = 1$ &
    \numprint{1004} & \numprint{15261} \\
  \bottomrule
\end{tabular}
\end{table}

\subsection{Discussion}

There is no benchmark parameter for which HLinear performs significantly worse than Flint. For random PLE matrices of size $100 \times 100$, classical Gaussian elimination in Flint is faster by 33\% than the one in HLinear. This strongly sustains our claim that competitive implementations of linear algebra can be written in functional programming languages.

We observe that HLinear performs vastly better than Flint when the number of columns grows. This is true even for the unoptimized fraction-free algorithm. In the most extreme test case, HLinear is faster by a factor of more than~34. Such a drastic improvement appears even more surprising, since HLinear cannot reorder or fuse elementary arithmetic operators. At the current stage, cache locality appears as the most likely reason for HLinear's performance boost.

\section{Conclusion}

We have demonstrated that the implementation of Gaussian elimination in a functional programming language can compete with C implementations and even outperform them. The design of our implementation was guided by the algebraic structure of intermediate steps. In particular, we exposed the iteration scheme of Gaussian elimination by unfolding explicitly a matrix to a vector of PLE~hooks. We believe that this feature made it possible for the compiler to rearrange them more easily while optimizing the code. Potential for such rearrangement is generally advertised as a fundamental advantage of functional programming, and our example shows how it comes into effect in a practical case.

Our fold-unfold implementation of Gaussian elimination is general enough to cover all division rings. Despite being very general it performs well in practice. With slight modification it can be extended to discrete valuation rings (e.g.\ the local ring~$\ZZ_p$). In a development version of HLinear, we have extended the PLE decomposition to an implementation of Hermite normal forms over Euclidean domains.

While for small matrices HLinear outperforms Flint, it falls behind for moderately sized ones. This is a hint towards insufficient strictness in our implementation. Preliminary experimentation with strictness, however, yielded worse performance on all scales. We will strive to introduce beneficial strictness in a future version of HLinear. Both block-based algorithms and multi-modular algorithms are compatible with the structure we have introduced. We will implement them in forthcoming work.

The splitting up into algebraically modeled intermediate steps also opens doors to formal verification. The remaining obstacle for this is the partially-defined nature of multiplication of PLE~hooks.

%%%%%%%%%%%%%%%%%%%%%%%%%%%%%%%%%%%%%%%%%%%%%%%%%%
%%% BIBLIOGRAPHY

\renewbibmacro{in:}{}
\renewcommand{\bibfont}{\normalfont\small\raggedright}
\renewcommand{\baselinestretch}{.8}

\Needspace*{4em}
%\begin{multicols}{2}
\printbibliography[heading=none]
%[heading=bibnumbered]
%\end{multicols}

% In case publishers don't support biblatex, switch to amsref (or plain bibtex)
% \bibliographystyle{alpha}
% \bibliography{bibliography.bib}

%%%%%%%%%%%%%%%%%%%%%%%%%%%%%%%%%%%%%%%%%%%%%%%%%%
%%% AFFILIATIONS

\addvspace{1em}
\titlerule[0.15em]\addvspace{0.5em}

\Needspace*{5em}
{\small
\setlength{\parskip}{1pt}

\noindent
Alexandru Ghitza

\noindent
School of Mathematics and Statistics,
University of Melbourne,
Parkville, VIC 3010, Australia

\noindent
E-mail: \url{aghitza@alum.mit.edu}

\noindent
Homepage: \url{http://aghitza.org}\\[1ex]

\Needspace*{5em}
\noindent
Martin Raum

\noindent
Chalmers tekniska högskola och G\"oteborgs Universitet,
Institutionen för Matematiska vetenskaper,
SE-412 96 Göteborg, Sweden

\noindent
E-mail: \url{martin@raum-brothers.eu}%

\noindent
Homepage: \url{http://raum-brothers.eu/martin}

}%\\[1.5ex]

\end{document}

%% vim: spell spelllang=en_us